\documentclass[12pt]{article}
\usepackage{graphicx} 
\begin{document}
\centerline{\bf Ising model $S=1$, 3/2 and $2$ on directed networks}

\bigskip
\centerline{F.W.S. Lima$^1$ and Edina M.S. Luz$^2$}
 
\bigskip
\noindent
$^1$Departamento de F\'{\i}sica, 
Universidade Federal do Piau\'{\i}, 64049-550, Teresina - PI, Brazil \\
$^2$Departamento de F\'{\i}sica, 
Universidade Estadual do Piau\'{\i}, 64002-150, Teresina - PI, Brazil 

\medskip
  e-mail:  wel@ufpi.br, edina@uespi.br
\bigskip
 
{\small Abstract: On {\it directed} Barab\'asi-Albert and Small-World networks the
 Ising model with spin $S=1$, $3/2$ and $2$ is now studied through Monte Carlo   simulations.
 In this model,  the order-disorder phase transition of the
 order parameter
 is well defined on Small-World networks for Ising model with spin $S=1$. We  calculate the value of the critical 
 temperature $T_{c}$ for several values of rewiring probability $p$  of the 
 {\it directed} Small-World network. This model on {\it directed} Small-World networks we
 obtained a second-order phase transition for $p=0.2$ and first-order
 phase   transition for $p=0.8$. The critical 
 exponentes $\beta/\nu$, $\gamma/\nu$ and
 $1/\nu$ were calculated for $p=0.2$. On {\it directed} Barab\'asi-Albert we
show that no there is phase transition for Ising model with spin $S=1$, $3/2$ 
and $2$. }
 
 
\bigskip

 {\bf Introduction}

 This paper deals with Ising spin on {\it directed} Barab\'asi-Albert(BA) and 
 Small-World(SW) networks. Sumour and Shabat \cite{sumour,sumourss} investigated Ising models with
 spin $S=1/2$ on {\it directed} BA networks \cite{ba} with
 the usual Glauber dynamics.  No spontaneous magnetisation was 
 found, in contrast to the case of {\it undirected}  BA networks
 \cite{alex,indekeu,bianconi} where a spontaneous magnetisation was
 found below a critical temperature which increases logarithmically with
 system size. In S=1/2 systems on {\it undirected}, scale-free hierarchical-lattice
 SW networks \cite{nihat}, conventional and algebraic
 (Berezinskii-Kosterlitz-Thouless) ordering, with finite transition
 temperatures, have been found. Lima and Stauffer \cite{lima} simulated
 {\it directed} square, cubic and hypercubic lattices in two to five dimensions
 with heat bath dynamics in order to separate the network effects  form
 the effects of directedness. They also compared different spin flip
 algorithms, including cluster flips \cite{wang}, for
 Ising-BA networks. They found a freezing-in of the 
 magnetisation similar to  \cite{sumour,sumourss}, following an Arrhenius
 law at least in low dimensions. This lack of a spontaneous magnetisation
 (in the usual sense)
 is consistent with the fact
 that if on a directed lattice a spin $S_j$ influences spin $S_i$, then
 spin $S_i$ in turn does not influence $S_j$, 
 and there may be no well-defined total energy. Thus, they show that for
 the same  scale-free networks, different algorithms give different
 results. The $q$-state Potts model has been studied in scale-free networks
 by Igloi and Turban \cite{igloi} and depending on the value of $q$ and the
 degree-exponent  $\gamma$ first- and second-order phase transitions
 are found, and also by Lima \cite{lima2} on {\it directed} 
 BA network, where only first-order phase transitions
 have being obtained independent of values of $q$ for values of 
 connectivity $z=2$ and $z=7$ of the {\it directed} BA network.
 More recently, Lima \cite{lima1} simulated the the Ising model
 for spin $S=1$ on {\it directed}
 BA network and different from the Ising model for
 spin $S=1/2$, the {\it unsual} order-disorder phase transition of 
 order parameter was seen; this effect is re-evaluated in the light of the time
depemdence presented below. We study the Ising model for spin $1$, $3/2$ and $2$ on {\it directed} 
 BA network. The Ising model with spin $1$, $3/2$ and $2$ was seen not to show  a {\it usual} spontaneous 
 magnetisation and this decay time for flipping of the magnetisation
 followed an  Arrhenius law for HeatBath algorithms that agree with the
 results of the Ising model for
 spin $S=1/2$ \cite{sumour,sumourss} on directed BA
 network.
  Edina and Lima \cite{lima3} calculate the exponentes
 $\beta/\nu$, $\gamma/\nu$, and $1/\nu$ exponents for
 majority-vote model on 
 {\it directed} SW networks of S\'anchez et al. \cite{sanches}, 
 and on these networks the exponents are 
 different from the Ising model a two-dimensional and independ on the values of
 rewiring probability $p$ of the 
 {\it directed} SW networks. Here we also study the Ising model for spin
 $S=1$  on {\it directed} SW, we obtained a second-order phase transition for 
$p=0.2$ and a first-order
 phase   transition for $p=0.8$. We also calculate the critical 
 exponents $\beta/\nu$, $\gamma/\nu$ and
 $1/\nu$ for $p=0.2$.

\bigskip

\begin{figure}[hbt]
\begin{center}
\includegraphics [angle=-90,scale=0.5]{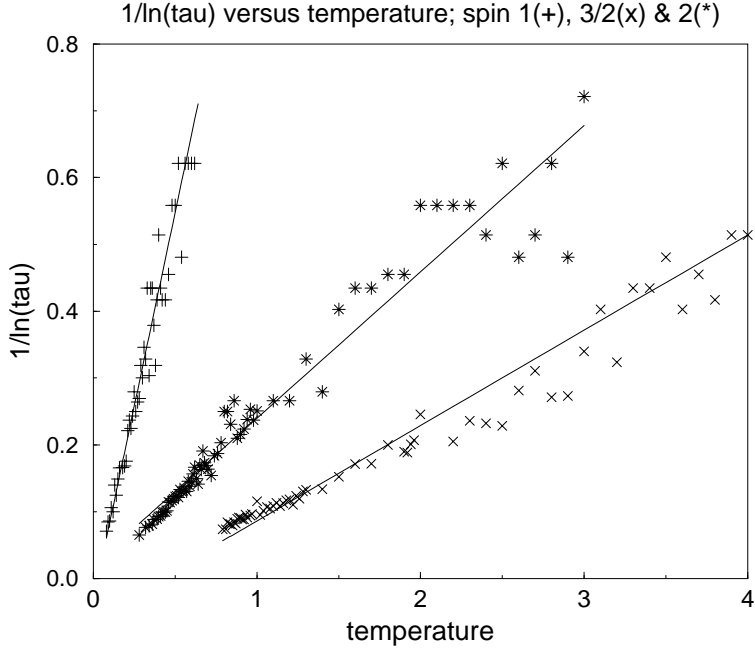}
\end{center}
\caption{Reciprocal logarithm of the relaxation times on directed BA networks for
$S=1$ to $S=2$.}
\end{figure}
 
\bigskip

\begin{figure}[hbt]
\begin{center}
\includegraphics [angle=-90,scale=0.5]{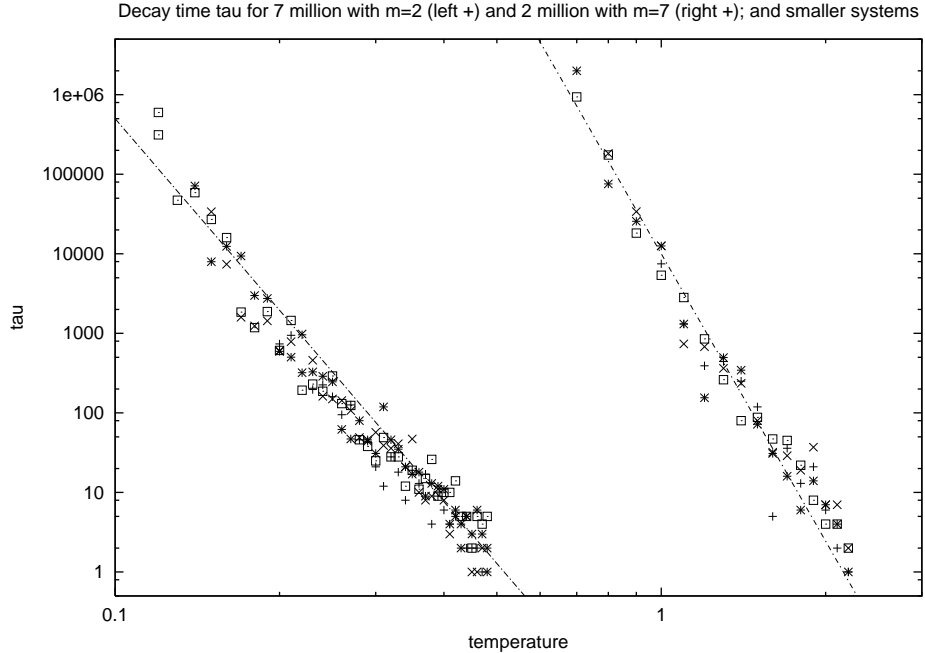}
\end{center}
\caption{Characteristic time for $M(\tau)=3/4$ using 7 million spins for
$m=2$ neighbours and $2$ million spins for $m=7$ neighbours (+). Ten,
hundred, and thousand times smaller systems are denoted by x, star, and squares.
We plot the median over nine samples in this log-log plot. The
two straight lines have negative slopes 8 (left) and  12 (right). From 
\cite{sumourss}.
}
\end{figure}
 
\bigskip

\begin{figure}[hbt]
\begin{center}
\includegraphics [angle=-90,scale=0.5]{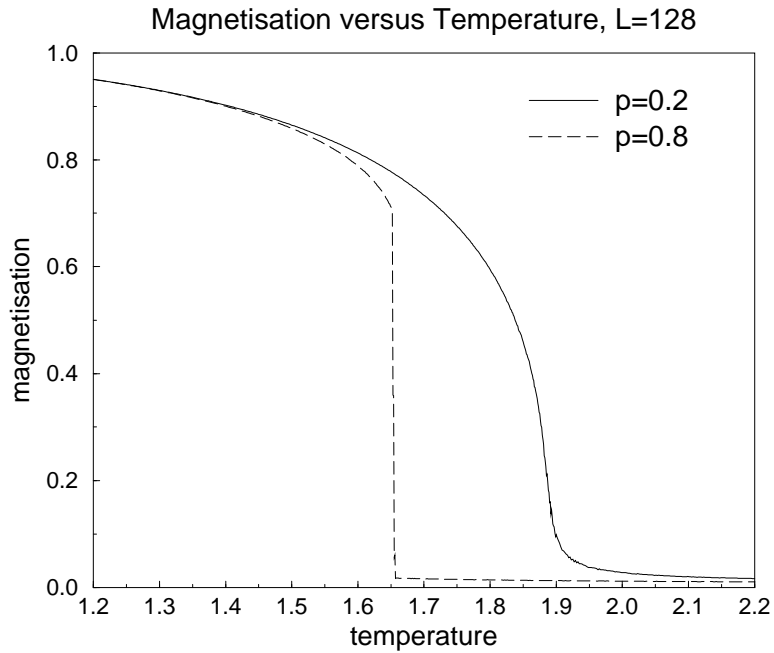}
\end{center}
\caption{
Magnetisation  as a function of the temperature $T$, for
$N=16384$ sites. A second-order phase transition for values of $p=0.2$ and a first-order phase   transition for $p=0.8$ .}
\end{figure}
 
\begin{figure}[hbt]
\begin{center}
\includegraphics[angle=-90,scale=0.29]{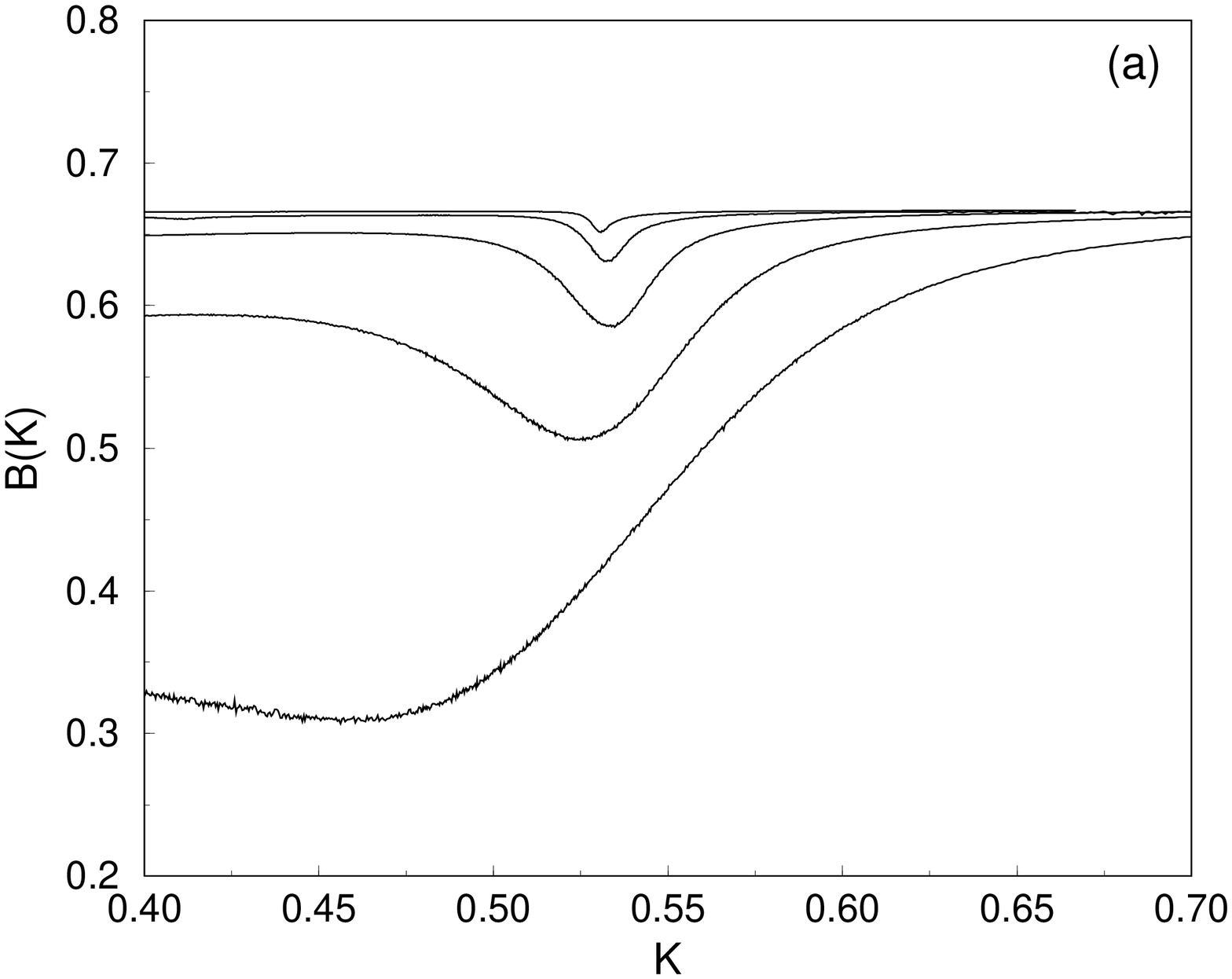}
\includegraphics[angle=-90,scale=0.29]{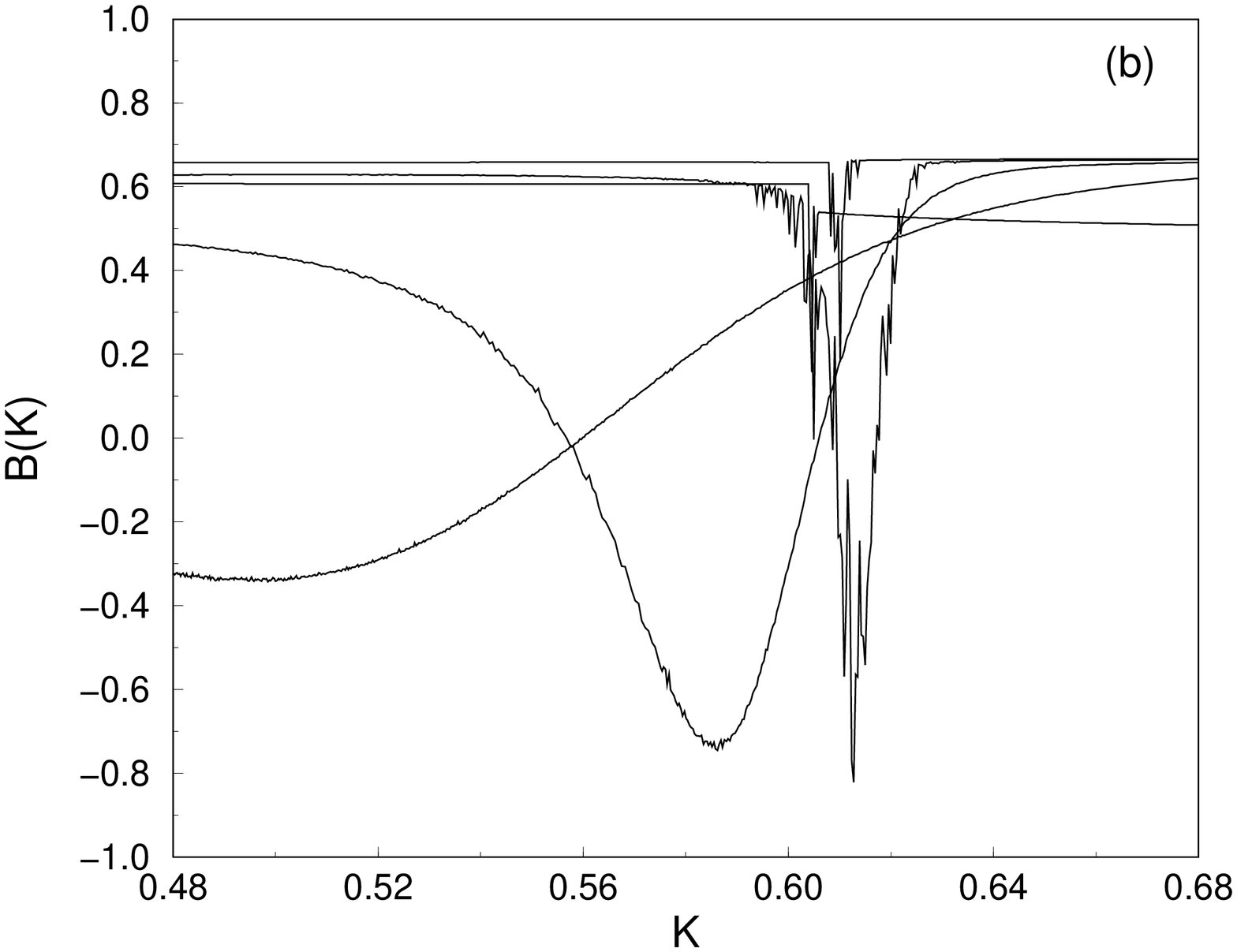}
\end{center}
\caption{
Binder's fourth-order cumulant as a function of $K$. We have $p=0.2$ (part a)
and $p=0.8$ (part b) for $L=8$, 16, 32, 64 and 128.}
\end{figure}
 
\begin{figure}[hbt]
\begin{center}
\includegraphics[angle=-90,scale=0.50]{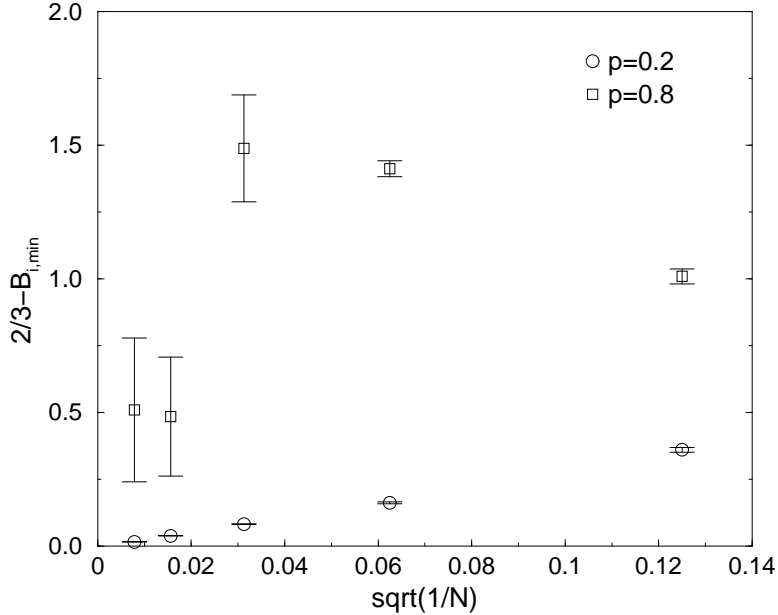}
\end{center}
\caption{$2/3-B_{i,min}$ for $p=0.2$ (circles) and $0.8$ (squares).
} 
\end{figure}
 
\begin{figure}[hbt]
\begin{center}
\includegraphics[angle=-90,scale=0.50]{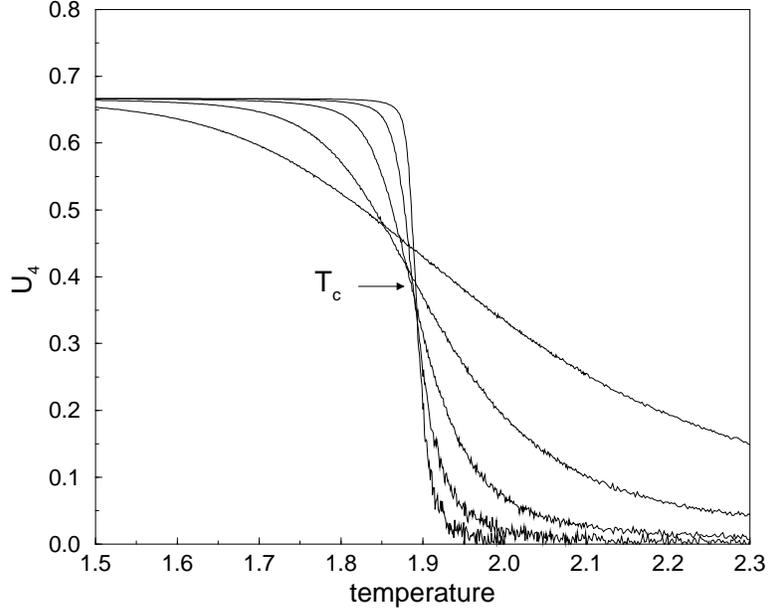}
\end{center}
\caption{Binder's fourth-order cumulant of the magnetisation versus temperature for $p=0.2$. }
\end{figure}

\bigskip

\begin{figure}[hbt]
\begin{center}
\includegraphics[angle=-90,scale=0.60]{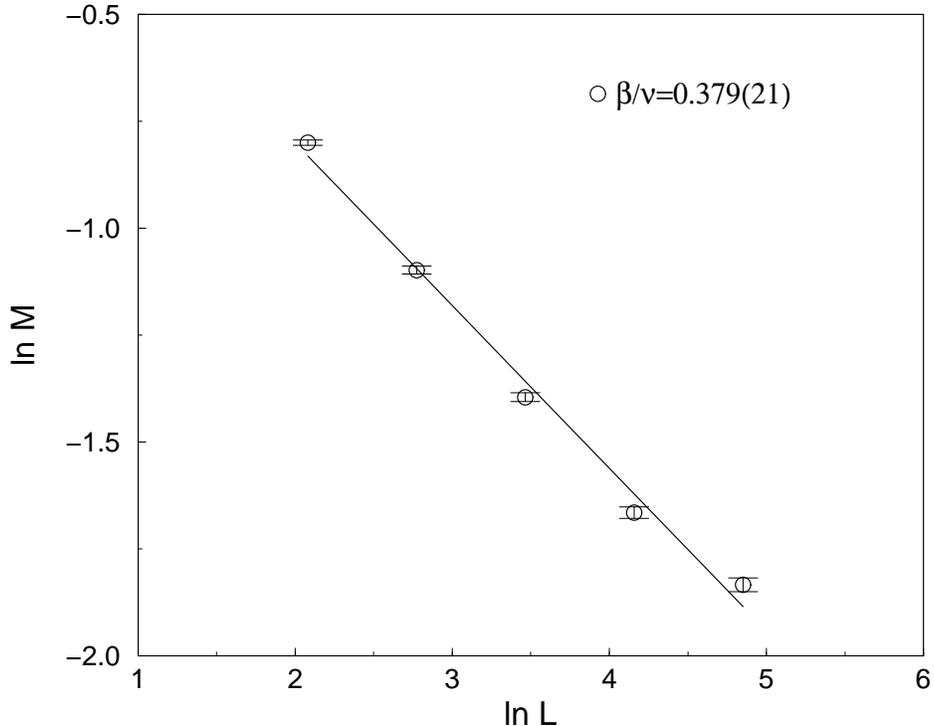}
\end{center}
\caption{ln $M(T_{c})$ versus  ln $L$ for $p=0.2$.}
\end{figure}
\bigskip
 
\begin{figure}[hbt]
\begin{center}
\includegraphics[angle=-90,scale=0.50]{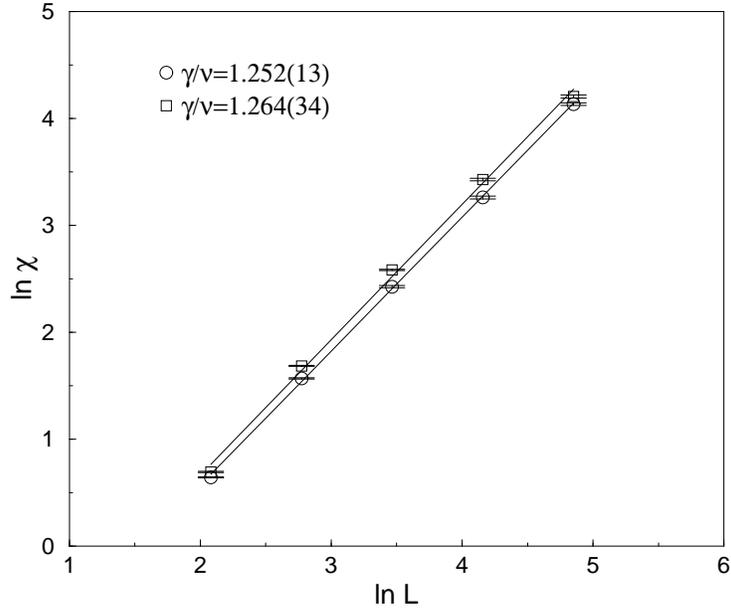}
\end{center}
\caption{  Plot of ln $\chi^{max}(L)$ (square) and ln$ \chi(T_{c})$ (circle) versus ln $L$ for $p=0.2$.}
\end{figure}

\begin{figure}[hbt]
\begin{center}
\includegraphics[angle=-90,scale=0.50]{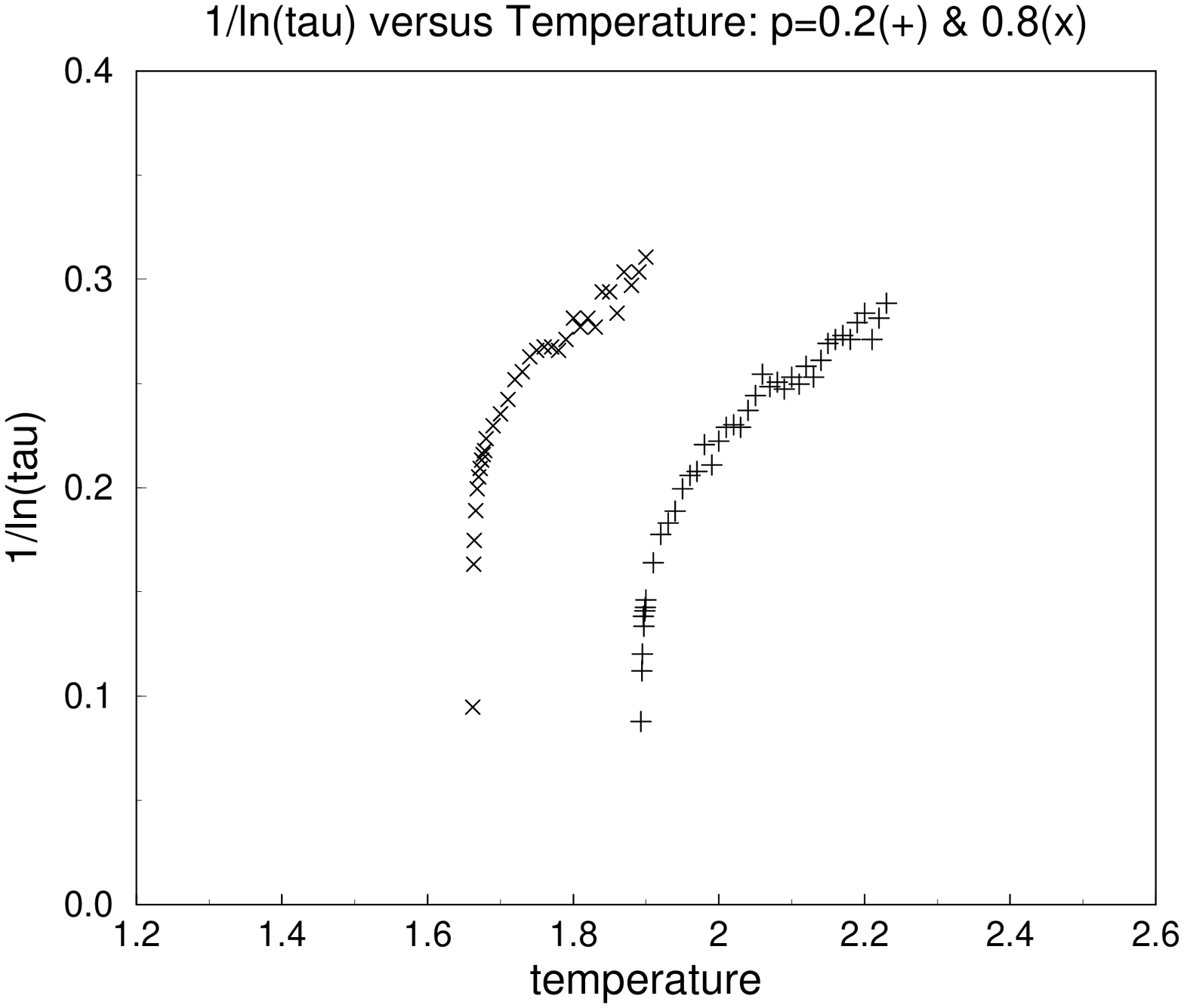}
\end{center}
\caption{ Reciprocal logarithm of the relaxation times on directed SW networks for versus $T$ for p=0.2(+) and $0.8$(x).}
\end{figure}

\bigskip

{\bf Model and Simulaton}

{\bf Ising model on {\it directed}
BA Networks:}

We consider the spin 1, 3/2 and 2 Ising models on {\it directed}
BA Networks, defined by a set of
spins variables ${S}$ taking the values  $\pm 1$ and
$0$ for $S=1$, $\pm 3/2$ and $1/2$ for $S=3/2$,  and $\pm 1$, $\pm 2$ and
$0$ for $2$, respectivaly,  located
on every site of a directed
BA Networks with $N$ sites.

The probability for spin $S_{i}$ change its state in this {\it directed}  networks is
\begin{equation}
%\begin{center}
p_{i}= 1/1+\exp(2E_{i}/k_BT),  \quad E_{i}=-J\sum_{k}S_{i}S_{k}
%\end{center}
\end{equation}
where $k$-sum runs over all nearest neighbors of $S_{i}$. In this network,
each new site
added to the network selects with connectivity $z=2$
already existing sites as neighbours influencing it; the newly
added spin does not influence these neighbours.

To study the spin $1$, $3/2$ and $2$ Ising models we start with
all spins up, a number of spins equal to $N=500000$, and Monte
Carlo step (MCS) time up $2,000,000$, in our simulations, one MCS is
accomplished
after all the $N$ spins are updated, here, with HeatBath Monte Carlo
algorithm. Then we vary the temperature and study nine samples.
The temperature is measured in units of critical temperature of the
square-lattice Ising model. We determine the time $\tau$ after which the
magnetisation has flipped its sign for the first time, and then take the
median value of our nine samples. So we get different values $\tau_{1}$
for different temperatures.

 In the study the critical behavior this Ising model (with spins 
$1$, $3/2$ and $2$) we define the variable 
$m=\sum_{i=1}^{N}S_{i}/N$ as normalized magnetisation.

Our BA simulations, using the HeatBath algorithm, indicate that the
spins $S=1$, $3/2$ and $2$ Ising model do not display a phase transition and
the plot of the time $1/\ln\tau$ versus temperature in Fig. 1 shows that our BA 
results for all spins agree with the modified Arrhenius law for relaxation time,
defined as the first time when the sign of the  magnetisation flips:$1/\ln(\tau)
\propto T +...$. This result agrees with the results of Sumour
et al. for spin $S=1/2$ \cite{sumourss} with $7$ million of sites, see Fig. 2(figure
retired of reference \cite{sumourss}) and our results are more reliable than 
both Lima \cite{lima1} and Sumour et al. \cite{sumuor2007} because that this 
are longer simulation times.
\bigskip

{\bf Ising model on {\it directed}
SW Networks:}

We consider the Ising model with spin $S=1$, on {\it directed} 
SW Networks, defined by a set of
spins variables ${S}$ taking the values $-1$, $0$ and
$+1$, situated on every site of a {\it directed} 
SW Networks with $N$ sites. 

The probability for spin $S_{i}$ to change its state in these {\it directed}
network is given eq. (1) where the 
sum runs over all nearest neighbors of $S_{i}$. In this network, created for 
S\'anchez et al. \cite{sanches}, we start from a two-dimensional square
lattice consisting of sites linked to their four nearest neighbors by both outgoing and
incoming links. Then, with probability $p$, we reconnect nearest-neighbor outgoing
links to a different site chosen at random. After repeating this process for every
link, we are left with a network with a density $p$ of SW {\it directed} links. Therefore,
with this procedure every site will have exactly four outgoing links and
varying (random) number of incoming links. To study the critical behavior of 
the model we use the HeatBath algorithm and define the variable $e=E/N$, where 
$E=2  \sum_{i}E_{i}$ and $m=\sum_{i=1}^{N}S_{i}/N.$ From the fluctuations of 
the $e$ measurements we can compute: the average $e$, 
the specific heat $C$ and the fourth-order cumulant of $e$,
\begin{equation}
 u(K)=[<E>]_{av}/N,
\end{equation}
\begin{equation}
 C(K)=K^{2}N[<e^{2}>-<e>^{2}]_{av},
\end{equation}
\begin{equation}
 B_{i}(K)=[1-\frac{<e^{4}>}{3<e^{2}>^{2}}]_{av},
\end{equation}
where $K=J/k_BT$, with $J=1$, and $k_B$ is the Boltzmann constant. 
Similarly, we can derive from the magnetization measurements
the average magnetization, the susceptibility, and the magnetic
cumulants,
\begin{equation}
 m(K)=[<|m|>]_{av},
\end{equation}
\begin{equation}
 \chi(K)=KN[<m^{2}>-<|m|>^{2}]_{av},
\end{equation}
\begin{equation}
 U_{4}(K)=[1-\frac{<m^{4}>}{3<|m|>^{2}}]_{av}.
\end{equation}
where $<...>$ stands for a thermodynamic average and $[...]_{av}$ square brackets
for a averages over the 20 realizations. 

In order to verify the order of the transition this model, we apply finite-size scaling (FSS) \cite{fss}. Intially we search for the minima of $e$ fourth-order parameter of  eq. (4). This quantity gives a qualitative as well as a quantitative description of the order of the transition \cite{mdk}. It is known \cite{janke} that this
parameter takes a minimum value $B_{i,min}$ at effective transition temperature 
$T_{c}(N)$. One can show \cite{kb} that for a second-order transition $\lim_{N\to \infty}$
$(2/3-B_{i,min})=0$, even at $T_{c}$, while at a first-order transition the same limit measurng the same quantity is small and $(2/3-B_{i,min})\neq0$.

A more quantitative analysis can be carried out through the FSS of the $C$ fluctuation 
$C_{max}$, the susceptibility maxima $\chi_{max}$ and the minima of the Binder parameter $B_{i,min}$. If the hypothesis of a first-order phase transition is correct, we should then expect, for large systems sizes, an asymptotics FSS behavior of the form
\cite{wj,pbc},
\begin{equation}
C_{max}=a_{C} + b_{C}N +...
\end{equation}
\begin{equation}
\chi_{max}=a_{\chi} + b_{\chi}N +...
\end{equation}
\begin{equation}
B_{i,min}=a_{B_{i}} + b{B_{i}}N +...
\end{equation}

Therefore, if the hypothesis of a second-order phase transition is correct, we should then expect, for large systems sizes, an asymptotics FSS behavior of the form
\begin{equation}
 C=C_{reg}+L^{\alpha/\nu}f_{C}(x)[1+...],
\end{equation}
\begin{equation}
 [<|m|>]_{av}=L^{-\beta/\nu}f_{m}(x)[1+...],
\end{equation}
\begin{equation}
 \chi=L^{\gamma/\nu}f_{\chi}(x)[1+...],
\end{equation}
\begin{equation}
\frac{dU_{4}}{dT}=L^{1/\nu}f_{U}(x)[1+...],
\end{equation}
where $C_{reg}$ is a regular background term, 
$\nu$, $\alpha$, $\beta$, and $\gamma$, are the usual critical
exponents, and $f_{i}(x)$ are FSS functions with
\begin{equation}
 x=(K-K_{c})L^{1/\nu}
\end{equation}
being the scaling variable, and the brackets $[1+...]$ indicate
corretions-to-scaling terms. Therefore, from the size dependence of $M$ and $\chi$
we obtained the exponents $\beta/\nu$ and $\gamma/\nu$, respectively.
The maximum value of susceptibility also scales as $L^{\gamma/\nu}$. Moreover, the
value of $T$ for which $\chi$ has a maximum, $ T_{c}^{\chi_{max}}=T_{c}(L)$,
is expected to scale with the system size as
\begin{equation}
T_{c}(L)=T_{c}+bL^{-1/\nu},
\end{equation}
were the constant $b$ is close to unity. Therefore, the  relations $(14)$ and $(16)$
are used to determine the exponent $1/\nu$.

We have performed Monte Carlo simulation on {\it directed} SW network with
various values of probability $p$. For a given $p$, we used systems
of size $L=8$, 16, 32, 64, and 128. We waited $50,000$ Monte Carlo
steps (MCS) to make the system reach the steady state, and the time averages were
estimated from the next $ 50,000$ MCS. In our simulations, one MCS is accomplished
after all the $N$ spins are updated. For all sets of parameters, we have generated
$20$ distinct networks, and have simulated $20$
independent runs for each distinct network.

\bigskip

{\bf Results and Discussion}

In Fig. 3 we show the dependence of the magnetisation $M$  on the temperature, obtained from simulations on {\it directed}
SW network with $L=128 \times 128$ sites and two values of probability $p$: $p=0.2$ for
second-order transition and $p=0.8$ for first-order transition. In Fig. 4 we plot
Binder's fourth-order cumulant of $e$ for different values of $L$ and two different values
of $p$: $p=0.2$ in part (a) and $p=0.8$ in part (b).
In Fig. 5 the difference $2/3-B_{i,min}$ is shown as a function of 
parameter $1/N$ for probability $p=0.2$ (circles) and $=0.8$ (squares) obtained 
from the data of Fig. 4. In Fig. 6 we show Binder's fourth-order cumulant of the magnetisation versus temperature for $p=0.2$. The temperature obtained is $T_{c}=1.890(6)$. Figs. 7  we plot the dependence of the magnetisation at $T=T_{c}$ versus the system
size $L$. The slopes of curves correspond to the exponent ratio $\beta/\nu$ of according
to Eq. (12).
The results show that the exponent ratio $\beta/\nu$  at $T_{c}$ is $0.379(21)$. 

In Fig. 8 we display the scalings for susceptibility at $T=T_{c}(L)$ , $\chi(T_{c}(L))$ (circles), and for its maximum amplitude, $\chi_{L}^{max}$ (squares), and the scalings for susceptibility at $T=T_{c}$ obtained from Binder's cumulant, $ \chi(T_{c})$ versus $L$ for probability $p=0.2$. The exponents ratio $\gamma/\nu$ are obtained from the slopes
of the straight lines. The exponents $\gamma/\nu$ of the two estimates agree (within errors). The values obtained are $\gamma/\nu=1.252(13)$ (circles) and $1.264(34)$ (squares), respectively. Therefore we can use the Eq. (16) , for $p=0.2$, obtain the critical exponent $1/\nu$, that is equal to $1.209(165)$ and $1.248(196)$ obtained of according
to Eq. (14). 
To improve our results obtained above we start with all spins up, a number of spins equal to $N=640000$, and time up 2,000,000 (in units of Monte Carlo steps per spins). Then we vary temperature $T$ and at each $T$ study the time dependence for 9 samples. We determine the time $\tau$ after which the magnetisation has flipped its sign for first time, and then take the median values of our nine samples. So we get different values $\tau_{1}$ for different temperatures $T$. In Fig. 9  show that the decay time goes to infinity at some positive $T$ value.
This behavior ensures that there is a phase transition for Ising model spin 
$S=1$ on {\it directed} SW network. 

\bigskip
 
{\bf Conclusion}
 
In conclusion, we have presented the Ising model spin spins $S=1$, $3/2$ and $2$ on
{\it directed} BA and {\it directed} SW network($S=1$). The  Ising model does not display a phase transition on {\it directed} BA for spins $S=1/2$, $1$, $3/2$ and $2$. In the {\it directed} SW network \cite{sanches} this model presents a
first- and second-order phase transition which occurs with 
probability $p=0.8$ and $0.2$, respectively. The exponents obtained for $p=0.2$ are different from the exponents the Ising model on square lattice, that it suggests that these exponents belong to one another class of universality. 

  The authors thanks  D. Stauffer for many suggestions and fruitful
discussions during the development this work and also for the revision of
this paper. We also acknowledge the Brazilian agency FAPEPI
(Teresina-Piau\'{\i}-Brasil) for  its financial support. This work also was supported the
system SGI Altix 1350 the computational park CENAPAD.UNICAMP-USP, SP-BRAZIL.

\end{document}